\documentclass[%
 reprint,
 amsmath,amssymb,
 aps,
]{revtex4-2}

\usepackage{graphicx}
\usepackage{dcolumn}
\usepackage{bm}
\usepackage{amsmath}
\usepackage{braket}
\usepackage{graphicx}
\usepackage{overpic}
\usepackage{lineno}
\usepackage{mathrsfs}
\usepackage{ragged2e}
\usepackage{subfigure}

\usepackage{ragged2e}

\makeatletter
\long\def\@makecaption#1#2{%
  \vskip\abovecaptionskip
  \sbox\@tempboxa{\footnotesize #1: #2}%
  \ifdim \wd\@tempboxa >\hsize
    \footnotesize\justifying #1: #2\par
  \else
    \global\@minipagefalse
    \hb@xt@\hsize{\footnotesize #1: #2\hfil}%
  \fi
  \vskip\belowcaptionskip
}
\makeatother

\begin{document}

\preprint{APS/123-QED}

\title{Time-Rescaled STIRAP Enables Compact Cascaded Frequency Conversion}

\hfill

\author{J. L. Montenegro Ferreira}

\email{lukas.montenegrof@gmail.com}

\affiliation{Instituto de Física, Universidade de São Paulo, 05315-970, Brazil}

\begin{abstract}
In this letter we explore how full frequency conversion can be performed in shorter optical devices by using a STIRAP-like protocol modified by the time rescaling shortcut to adiabaticity. We first show a detailed, step-by-step process of how the coupled equations for two simultaneous three-wave mixing processes can be written in terms of a STIRAP-like system. From the properties of the STIRAP protocol, we know that the ensuing process will result in robust conversion, but require long propagation distances inside the bulk crystal or waveguide. We then apply the time rescaling shortcut to adiabaticty in the conversion process to create a TR-STIRAP protocol, showing that full and robust conversion is also obtained, but at a fraction of the propagation distance. We also show how the original shaping of the coupling coefficients required by the TR-STIRAP can be approximated by Gaussian functions with high conversion fidelity, thus simplifying the experimental implementation. Finally, a possible application of the new protocol is discussed in the context of single-photon detection in the MIR.
\end{abstract}

\maketitle

\section{Introduction}

Reliable and efficient frequency conversion that creates new and not directly available frequencies from an initial source is an useful technique in several areas of optics and photonics, such as wave generation in the terahertz range \cite{haitao}, optical multiplexing and de-multiplexing \cite{qiang,thet}, white laser generation \cite{Li} and single photon detection \cite{Ma}. This conversion could in principle be achieved through nonlinear phenomena such as three-wave mixing (TWM), although a single process could achieve low conversion efficiency due to phase-matching conditions. An interesting alternative is to use multi- and simultaneous TWM processes such as cascaded configurations where, for example, an initial signal is first converted to its second harmonic (SH), then converted again through sum or difference frequency generation to the desired frequency, much in the same way that electrons can be excited from a ground state, then spontaneously decay to new levels in atomic transitions \cite{shankar,paul}. However, a problem with this approach is that part of the signal's energy can be permanently lost to the intermediary field, also curtailing the overall conversion efficiency.

A solution to this problem can be found once we further exploit the analogy between nonlinear cascaded conversion and three-level atomic transitions: it allows us to use a well-known technique called \textit{Stimulated Raman Adiabatic Passage} (STIRAP), which guarantees that, if the coupling coefficients between energy levels vary sufficiently slowly in time, full transition between the first and third levels can occur without populating the second one \cite{bergmann,vitanov}. Therefore, a STIRAP-like frequency conversion can fully convert the original input to the desired output frequency, without populating the intermediary field, if the coupling coefficients between different frequencies vary sufficiently slowly upon propagation inside the nonlinear material \cite{yong,liu,pragati,dou,quan}.

A final set of problems arise from the fact that a STIRAP-like conversion process requires a long nonlinear medium (being it a bulk crystal or waveguide), which can result in optical losses related to scattering or slightly non-collinear beams. Their size also makes them harder to integrate into complex photonic devices. A proposed solution for these problems is to explore the idea of shortcuts to adiabaticity (STAs): processes that harness the same advantages of adiabatic processes (like STIRAP) while operating at shorter time intervals (propagation distances in our case) \cite{takuya,Xia,You,Sheng}. From the several STAs available in the literature, e.g. counterdiabatic method, fast-forward technique, reverse engineering, etc. \cite{odelin}, we choose in this letter to use the time rescaling (TR) method to explore how to perform cascaded nonlinear frequency conversion through a shortcut version of the STIRAP protocol. This choice is motivated by the fact that (i) this STA was proposed fairly recently and still requires exploration in other physical systems and (ii) due to its inherent advantages when compared to other STAs, such as not requiring additional coupling fields, theoretical simplicity, good robustness and control, and the fact that it reproduces the same evolution path along the Hilbert space as that of the original protocol \cite{bertulio, lukas}.

In the next section, we show how the coupled wave equations for the fields involved in the cascaded process can be manipulated to yield STIRAP-like dynamics. The following section employs the TR method and shows how the conversion process can be performed in considerably shorter distances. In section 4 we explore the fidelity of the method and how it changes if we make some simplifying modifications in the TR coupling coefficients. Section 5 discusses a concrete example, in which the method is applied in single-photon detection. We finish with a general discussion of the results and an exploration of future perspectives.

\section{STIRAP-like Conversion Dynamics}\label{stirap}

\hfill

\begin{figure}[h!]
    \centering
    \includegraphics[width=1.0\linewidth]{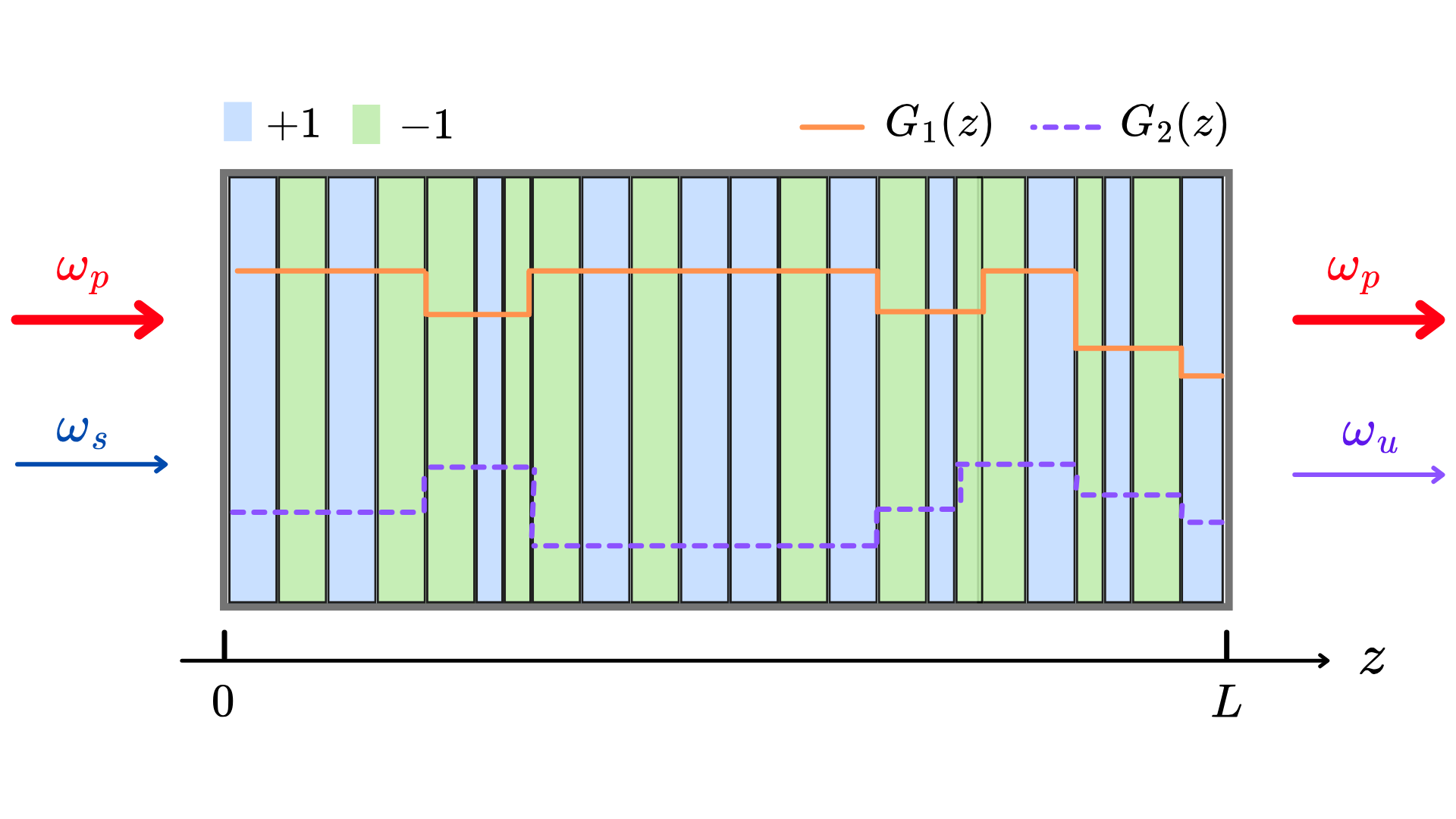}
    \caption{Conceptual representation of the cascaded conversion process. A seed field $\omega_s$ interacts with a strong pump field $\omega_p$ inside a nonlinear crystal with two different effective susceptibilities, resulting in a cascaded SFG process. The cumulative result is to obtain the up-converted field $\omega_u$ without populating the intermediary field $\omega_i$. The alternating color patterns in the crystal represent regions of opposite polling. This binary pattern can be described in terms of spatial Fourier components, two of which correspond to the desired effective susceptibility functions.}
    \label{ref}
\end{figure}

We begin by considering the cascaded frequency conversion system shown in Fig. \ref{ref}. An initial field with frequency $\omega_s$, which we call `seed', interacts with a strong pump field ($\omega_p$) inside a second order nonlinear crystal of size $L$ whose polling pattern gives rise to two effective nonlinear susceptibility components $G_1(z)$ and $G_2(z)$. The first component is such that it allows for sum frequency conversion (SFQ) with quasi phase-matching (QPM) between the two fields, generating a third field at frequency $\omega_i = \omega_s + \omega_p$, which we call the `intermediary'. The second component then allows the pump to interact with the new generated frequency, resulting in a new SFG with QPM process that creates the final desired field with frequency $\omega_u = \omega_i + \omega_p$, which we call `up-converted' . Although there are many possible configurations for this cascaded system, depending on the desired process, our particular choice is motivated by the fact that (i) it allows easy manipulation of the coupling coefficients to be presented later and (ii) it has a nice correspondence to the particular experimental process that we shall address in section 5. 

If the fields are taken as quasi-monochromatic, they can be written as 
\begin{equation}
    E_{j}(\vec{r},\omega) = A_{j}(z)\,e^{i(k_{j}z-\omega_{j}t)} + c.c.,
\end{equation}
where $A_{j}(z)$ is the field amplitude and $k_{j}$ the wave-number, $j = s,i,u,p$. For the purposes of this work, we approximate these fields to plane waves, so that there is no transverse spatial distribution. However, the field can vary with propagation as a result of the conversion processes, justifying $A_j = A_j(z)$. An exception is made for the pump field, which we consider to be real and strong enough to be in the undepleted regime. Under the slowly-varying envelope approximation (SVEA), the equations of motion for these different fields are given by

\begin{subequations}
\begin{align}
    i\frac{\partial A_{s}}{\partial z}
    &= \frac{\chi^{(2)}_{1}(z)\,\omega_{s}}{n_{s}c}
    A^{*}_{p}A_{i}\,e^{i\Delta k_{1}z},
    \label{field1}\\
    i\frac{\partial A_{i}}{\partial z}
    &= \frac{\chi^{(2)}_{1}(z)\,\omega_{i}}{n_{i}c}
    A_{p}A_{s}\,e^{-i\Delta k_{1}z}
    +\frac{\chi^{(2)}_{2}(z)\,\omega_{i}}{n_{i}c}
    A^{*}_{p}A_{u}\,e^{-i\Delta k_{2}z},
    \label{field2}\\
    i\frac{\partial A_{u}}{\partial z}
    &= \frac{\chi^{(2)}_{2}(z)\,\omega_{u}}{n_{u}c}
    A_{p}A_{i}\,e^{i\Delta k_{2}z}.
    \label{field3}
\end{align}
\end{subequations}
\\
where $n_j$ is the refractive index of the material in that frequency, $c$ is the speed of light in vacuum and $\Delta k_{1} = k_{p}+k_{s}-k_{i} + 2\pi/\Lambda_1$, $\Delta k_{2} = k_{u}-k_{p}-k_{i} + 2\pi/\Lambda_2$ are the phase mismatches, $\Lambda_1,\Lambda_2$ being the polling distances associated with each susceptibility component.

\hfill

The function $\chi^{(2)}_{1,2}(z)$  is the second order nonlinear susceptibility of the material. While this parameter is traditionally constant along the propagation coordinate $z$, we can harness the ferroeletric properties present in certain crystals used for second order nonlinear processes (e.g. LiNbO$_3$ and KTiOPO$_4$) to cause local domain inversions on the sign of the susceptibility, thus creating a binary pattern shown in Fig. \ref{ref}. If the proper pattern is engraved, the expansion of the function will result in the two spatial Fourier components of interest,
\\
\begin{equation}
\begin{split}
        \chi^{(2)}(z) = G_{1}(z)e^{2i\pi/\Lambda_1} + G_{2}(z)e^{2i\pi/\Lambda_2} + ...\\
    =\chi^{(2)}_{1}(z)e^{2i\pi/\Lambda_1} + \chi^{(2)}_{2}(z)e^{2i\pi/\Lambda_2} + ...
\end{split}
\end{equation}
\\
Details on how to choose an adequate pattern or how to experimentally perform ferroelectric domain inversion can be found in \cite{arie,tolga}.

\hfill

To approximate the equations of motion to those of a STIRAP-like system, we make the following changes in variables: $A_{s}(z) \rightarrow \tilde{A}_{s}(z) = A_{s}(z)$, $A_{i}(z) \rightarrow \tilde{A}_{i}(z) = A_{i}(z)e^{i\Delta k_{1}z}$ and $A_{u}(z) \rightarrow \tilde{A}_{u}(z) =  A_{u}(z)e^{i(\Delta k_{1}-\Delta k_{2})z}$.
Applying these in eqs. \eqref{field1}-\eqref{field3} and rearranging, we arrive at equations of motion:

\begin{widetext}
\begin{subequations}
\begin{align}
    i\frac{\partial \tilde{A}_{s}(z)}{\partial z}
    &= \frac{\chi^{(2)}_{1}(z)\,\omega_{s}}{n_{s}c}
    A^{*}_{p}\tilde{A}_{i}(z),
    \label{mot1}\\
    i\frac{\partial \tilde{A}_{i}(z)}{\partial z}
    &= \frac{\chi^{(2)}_{1}(z)\,\omega_{i}}{n_{i}c}
    A_{p}\tilde{A}_{s}(z) + \Delta\tilde{A}_{i}(z)
    +\frac{\chi^{(2)}_{2}(z)\,\omega_{i}}{n_{i}c}
    A^{*}_{p}\tilde{A}_{u}(z),
    \label{mot2}\\
    i\frac{\partial \tilde{A}_{u}(z)}{\partial z}
    &= \frac{\chi^{(2)}_{2}(z)\,\omega_{u}}{n_{u}c}
    A_{p}\tilde{A}_{i}(z) + \delta\tilde{A}_{u}(z).
    \label{mot3}
\end{align}
\end{subequations}
\end{widetext}

where the $\Delta = k_i - (k_s+k_p+2\pi/\Lambda_1) $ and $\delta = [k_{u} - (k_i+k_p+2\pi/\Lambda_2)] - [k_i - (k_s + k_p + 2\pi/\Lambda_1)]$ phase mismatches are analogous versions of the single-photon and two-photon detunings in STIRAP literature \cite{vitanov,shore}.

\hfill

We now proceed to quantization. Using the canonical approach, we exchange the amplitudes of each field in eqs. \eqref{mot1}-\eqref{mot3} by annihilation operators
\begin{equation}
    \tilde{A}_{j} \rightarrow \alpha_{j}\hat{a}_{j}(\omega_{j}),
\end{equation}
\\
where $\alpha_j$ is a normalization constant given by $\alpha_{j} = \sqrt{2\hbar \omega_{j}/n_{j}\epsilon_{0}c}$ \cite{bier}, $\hbar$ being the reduced Planck constant and $\epsilon_0$ the electric permittivity of vacuum. These operators obey the canonical commutation rules given by
\begin{subequations}
\begin{align}
    [\hat{a}_{i},\hat{a}^{\dagger}_{j}] &
    = \delta_{ij},
    \label{com1}\\
   [\hat{a}_{i},\hat{a}_{j}] 
   &= 0,
    \label{com2}\\
    [\hat{a}^{\dagger}_{i},\hat{a}^{\dagger}_{j}] 
    &= 0.
    \label{com3}
\end{align}
\end{subequations}
\\
After quantizing the fields, we re-organize eqs. \eqref{mot1}-\eqref{mot3} to obtain 
\\
\begin{subequations}
\begin{align}
    i\frac{\partial\, \hat{a}_{s}(z)}{\partial z}
    &= \kappa_{1}(z)
    \hat{a}_{i}(z),
    \label{fin1}\\
    i\frac{\partial\, \hat{a}_{i}(z)}{\partial z}
    &= \kappa_1(z)\hat{a}_{s}(z) + \Delta\hat{a}_{i}(z)
    +\kappa_{2}(z)\hat{a}_{u}(z),
    \label{fin2}\\
    i\frac{\partial\, \hat{a}_{u}(z)}{\partial z}
    &= \kappa_2(z)\hat{a}_{i}(z) + \delta\hat{a}_{u}(z),
    \label{fin3}
\end{align}
\end{subequations}
\\
where the coupling coefficients $\kappa_{1,2}(z)$ are given by
\\
\begin{subequations}
\begin{align}
    \kappa_1(z)
    &= \sqrt{\frac{2\omega_s\omega_iP_{p}}{\epsilon_0 n_sn_in_pc^3A}}\chi^{(2)}_{1}(z),
    \label{coeff1}\\
   \kappa_2(z)
    &= \sqrt{\frac{2\omega_i\omega_uP_{p}}{\epsilon_0 n_in_un_pc^3A}}\chi^{(2)}_{2}(z),
    \label{coeff2}
\end{align}
\end{subequations}
\\
$P_{p}$ being the power of the pump field and $A$ its transverse beam area. Note that by taking the pump field as undepleted, we were able to linearize the dynamics of our system, adding one of the field amplitudes as a constant inside the coupling coefficients. This procedure is valid here because all other fields are incredibly weak, populated by at most a single photon at any point along $z$. While the STIRAP dynamics can also be applied to the frequency conversion of classical fields, in a scenario where the seed field has an intensity comparable to the pump, the latter begins to be depleted and the dynamics has now to be described in terms of a four-field system, requiring a complete modification of our framework.

\hfill

The system described by eqs. \eqref{fin1}-\eqref{fin3} is already analogous to the three-level system where STIRAP occurs, if we take $\delta=0$. We choose to perform a final step, which is to change from the Heisenberg to the Schrodinger picture. Here, the three level system is described in terms of the basis vectors $\ket{1} = \ket{1_s,0_i,0_u}$, $\ket{2} = \ket{0_s,1_i,0_u}$, $\ket{3} = \ket{0_s,0_i,1_u}$, which represent one photon in each of the fields of interest, while the other two have no occupation. Since the photon arrives with frequency $\omega_s$, the initial state of the system is $\ket{1}$, which can then vary as this photon interacts within the material. The general state of the field at any point in $z$ will therefore be given by
\\
\begin{equation}\label{state}
    \ket{\psi(z)} = \sum^{3}_{n=1} c_{n}(z)\ket{n}, 
\end{equation}
\\
where $c_{j}(z)$ are the amplitudes associated with the probability of measuring the photon with a given frequency at $z$. Applying eq.\eqref{state} in eqs. \eqref{fin1}-\eqref{fin3} yields
\\
\begin{subequations}
\begin{align}
    i\frac{\partial\; c_{1}(z)}{\partial z}
    &= \kappa_{1}(z)
    c_{2}(z),
    \label{coef1}\\
    i\frac{\partial\,c_{2}(z)}{\partial z}
    &= \kappa_1(z)c_{1}(z) + \Delta c_{2}(z)
    +\kappa_{2}(z)c_{3}(z),
    \label{coef2}\\
    i\frac{\partial\, c_{3}(z)}{\partial z}
    &= \kappa_2(z)c_{2}(z), 
\end{align}
\end{subequations}
\\
which can be re-written as
\\
\begin{equation}\label{dynamics}
    i\partial_{z}\ket{\psi(z)} = \mathcal{H}_{eff}\ket{\psi(z)},
\end{equation}
\\
with the effective Hamiltonian $\mathcal{H}_{eff}$
\begin{equation}\label{stirap-dy}
\mathcal{H}_{eff}(z)
    = 
    \begin{pmatrix}
        0 & \kappa_{1}(z) & 0\\
        \kappa_{1}(z) & \Delta & \kappa_{2}(z)\\
        0 & \kappa_{2}(z) & 0
    \end{pmatrix}.
\end{equation}
\\
This is precisely the STIRAP dynamics as shown in \cite{vitanov, shore}, with the exception that the system evolves not in time, but by propagating in the nonlinear material.

\hfill

The coupling coefficients in eqs.  \eqref{coeff1} and \eqref{coeff2} are analogous to the Rabi frequencies of the pulses promoting population inversion between the atomic levels in traditional STIRAP. Note that while the susceptibility function itself has an upper limit depending on the properties of the crystal (typically between a few to tens of pm/V), we can increase or decrease these coefficients at will by simply changing the amount of pump power sent through the crystal. 

\hfill

One of the eigenstates of the Hamiltonian in eq. \eqref{stirap-dy} is given by
\begin{equation}
        \ket{n_{0}(z)} = \cos{[\theta(z)]}\ket{1} - \sin{[\theta(z])}\ket{3},
\end{equation}
with $\theta(z) = \arctan{[\kappa_{1}(z)/\kappa_{2}(z)]}$ being the mixing angle. This is known as a dark state \cite{celso} and as long as the system is kept in this state, it is trapped between $\ket{1}$ and $\ket{3}$, without populating $\ket{2}$ at any point in $z$. At $z=0$, the initial state of the system is $\ket{\psi_{i}} = \ket{1} \equiv \ket{n_{0}(\theta=0)}$. Then, to perform total conversion, we simply need to change the coupling coefficients along $z$ so that at $z=L$, $\theta = \pi/2$ and $\ket{\psi_{f}} = \ket{3}$. This requires that $\kappa_{1}(0) \ll \kappa_{2}(0)$ and $\kappa_{1}(L) \gg \kappa_{2}(L)$, the well known condition for STIRAP called `counterintuitive' ordering. Furthermore, since STIRAP is an adiabatic process, in order to keep the system at $\ket{n_{0}(z)}$, the rate at which the mixing angle changes with propagation must be sufficiently small. This condition can be expressed as 
\begin{equation}\label{ad-con}
    \left|\frac{\partial\theta}{\partial z}\right| \ll \kappa(z),
\end{equation}
with $\kappa(z) = \sqrt{\kappa^{2}_{1}(z)+\kappa^{2}_{2}(z)}$ \cite{shore}.

\hfill

\hfill

\begin{figure}[ht]
    \centering
    \begin{minipage}{0.80\columnwidth}
        \centering
        \begin{overpic}[width=\linewidth]{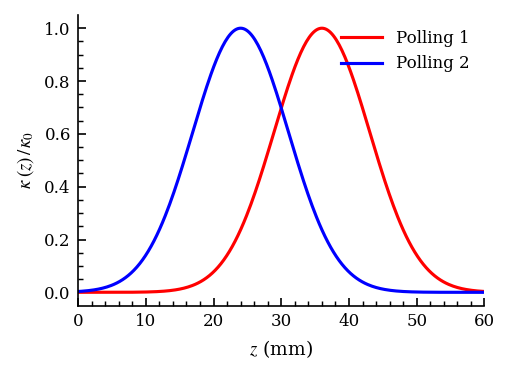}
            \put(0,70){\makebox(0,0){\footnotesize (a)}}
        \label{original1}
        \end{overpic}
    \end{minipage}
    \vfill
    \begin{minipage}{0.80\columnwidth}
        \centering
        \begin{overpic}[width=\linewidth]{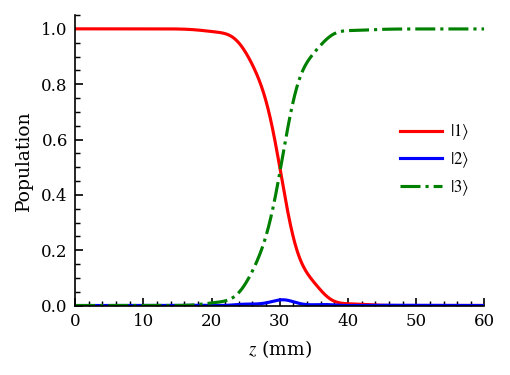}
            \put(0,70){\makebox(0,0){\footnotesize (b)}}
        \label{original2}
        \end{overpic}
    \end{minipage}
    \caption{(a) Gaussian coupling coefficients created by engineering
    the nonlinear susceptibility of the crystal. Here we used
    $L = 60$ mm, $d = L/10$ and $\sigma = L/5$.
    (b) Populations of the three states of the field as a function
    of the propagation distance.}
\end{figure}

To exemplify the STIRAP-like conversion dynamics, we choose Gaussian functions as coupling coefficients, in order to maintain the analogy with the Gaussian pulses used in atomic population inversion,
\\
\begin{subequations}
\begin{equation}\label{gauss1}
    \kappa_{1}(z) = \kappa_{0}\;e^{-(z-L/2-d)^{2}/s^{2}},
\end{equation}
\begin{equation}\label{gauss2}
    \kappa_{2}(z) = \kappa_{0}\;e^{-(z-L/2+d)^{2}/s^{2}},
\end{equation}
\end{subequations}
\\
with $2d$ being the separation between the Gaussians peaks and $s$ their width (see Fig. \ref{original1} a.). We choose to work with $\Delta=0$ for simplicity, but the STIRAP protocol is known to be robust for $\Delta \neq 0$. For our particular simulations, we choose $d = L/10$ and $s = L/5$ in order to make sure that (i) both Gaussians have an area $> 99\%$ within the crystal and (ii) the normalized overlap between the two Gaussians is $< 0.9$, since the STIRAP protocol fails in a full overlap situation \cite{vitanov}. The peak of these Gaussian functions is given by $\kappa_0$ and depends of several parameters as shown in eqs. \eqref{coeff1}-\eqref{coeff2}. Typical values in previous works \cite{pragati,porat} have $\kappa_0$ in the $10^{-1}-10^{1}$ cm$^{-1}$ region. Picking $\kappa_0 = 1$ cm$^{-1}$ as a reasonable value, we can evaluate eq. \eqref{ad-con} and choose $L=60$ mm as the size of our crystal in order to maintain the system comfortably within the adiabatic regime.

Using eqs. \eqref{gauss1} and \eqref{gauss2} for the coupling coefficients, we can solve eq. \eqref{dynamics} numerically and obtain the populations of the different states along $z$, as shown in Fig. \ref{original2} b. If the system remains in the adiabatic regime, we should expect
\\
\begin{subequations}
\begin{align}
    P_1(z) &=
    |\braket{1|\psi(z)}|^2 \approx |\braket{1|n_0(z)}|^2 = \cos^{2}[\theta(z)] ,
    \label{prob1}\\
    P_2(z) &=
    |\braket{2|\psi(z)}|^2 \approx |\braket{2|n_0(z)}|^2 = 0,
    \label{prob2}\\
    P_3(z) &=
    |\braket{3|\psi(z)}|^2 \approx |\braket{3|n_0(z)}|^2 = \sin^{2}[\theta(z)], 
    \label{prob3}
\end{align}
\end{subequations}
\\
which seems compatible with the results obtained.

\section{Shortening the Conversion Process}

To shorten the length necessary to perform full conversion, we may apply the concept of shortcuts to adiabaticity (STAs): a family of techniques that instruct how to modify adiabatic protocols so that they retain their same robust control and fidelity, but are performed in shorter times. Building on the known analogy between quantum states that evolve in time via the Schrödinger equation and the evolution of electromagnetic fields upon propagation via the paraxial wave equation \cite{erick}, we can adapt these techniques to work with the frequency conversion process. For the present work, the chosen STA is the time rescaling (TR) method \cite{bertulio}, where the evolution variable (in our case, the propagation coordinate $z$) is re-parametrized
\\
\begin{equation}
    z \rightarrow f(z),
\end{equation}
\\
and the system evolves according to a new Hamiltonian given by
\\
\begin{equation}
    \mathcal{H}(z) \rightarrow \mathscr{H}(z) = \mathcal{H}[f(z)]\frac{\partial\,f(z)}{\partial z},
\end{equation}
\\
so that if the original evolution occurs between $z_i$ and $z_{f}$, the new evolution occurs between $z'_{i} = f^{-1}(z_{i})$ and $z'_{f} = f^{-1}(z_{f})$. As shown in \cite{bertulio}, this protocol will represent a shortcut to adiabaticity if we choose a rescaling function $f(z)$ that satisfies the following conditions: (i) $\mathscr{H}(0) = \mathcal{H}(0)$, (ii) $\mathscr{H}(z_{f}) = \mathcal{H}(z'_{f})$, (iii) $z_{i} = z'_{i}$ and (iv) $z_{f} > z'_{f}$. One such function is
\\
\begin{equation}
    f(z) = az - \frac{z_{f}}{2\pi a}(a-1)\sin{\left(\frac{2\pi a}{z_{f}}z\right)},
\end{equation}
\\
and we can easily check that $z'_{f} = z_{f}/a$, making the protocol shorter as long as $a > 1$. Under this last condition, $a$ is called the contraction parameter. Since $z_{f} = L$, we can in principle reduce the size of the crystal as much as we want by conveniently choosing the value of $a$. 

\hfill

\begin{figure}[h!]
    \centering
\begin{minipage}{0.80\columnwidth}
        \centering
        \begin{overpic}[width=\linewidth]{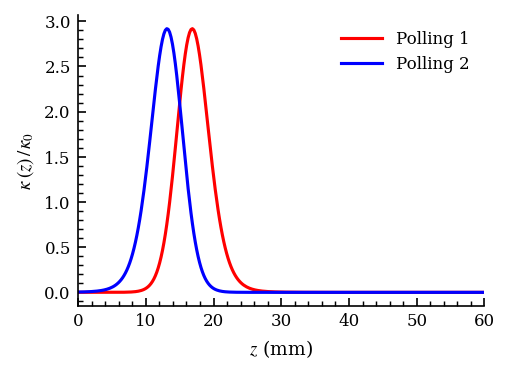}
            \put(0,70){\makebox(0,0){\footnotesize (a)}}
        \end{overpic}
    \end{minipage}
    \vfill
    \begin{minipage}{0.80\columnwidth}
        \centering
        \begin{overpic}[width=\linewidth]{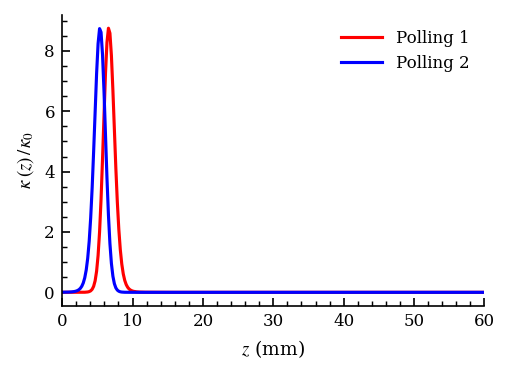}
            \put(0,70){\makebox(0,0){\footnotesize (b)}}
        \end{overpic}
    \end{minipage}
    \vfill
     \begin{minipage}{0.80\columnwidth}
        \centering
        \begin{overpic}[width=\linewidth]{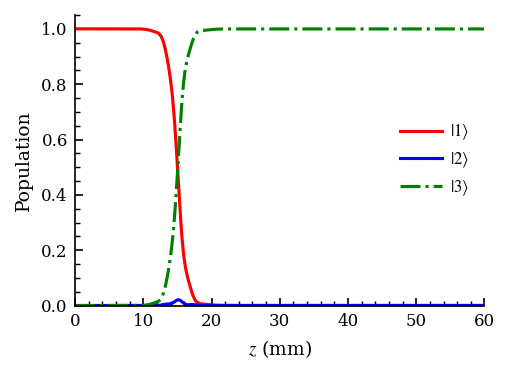}
            \put(0,70){\makebox(0,0){\footnotesize (c)}}
        \end{overpic}
    \end{minipage}
    \vfill
 \begin{minipage}{0.80\columnwidth}
        \centering
        \begin{overpic}[width=\linewidth]{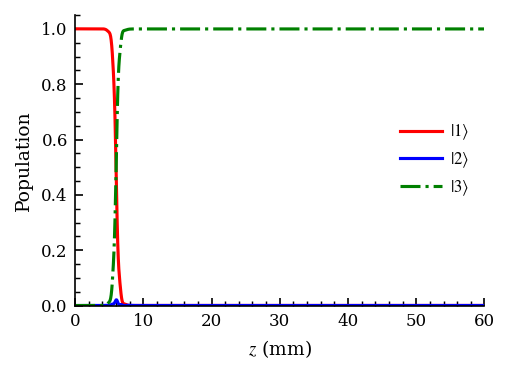}
            \put(0,70){\makebox(0,0){\footnotesize (d)}}
        \end{overpic}
    \end{minipage}
    \caption{Gaussian coupling coefficients under the TR-conversion process for (a) $a = 2.0$ (b) and $a=5$. (c) Populations of the three states for $a=2.0$ and (d) $a = 5.0$. All the other parameters were kept the same as in Fig. \ref{original1}.}
    \label{new}
\end{figure}

Returning to the example of the previous section, we simply substitute the new Hamiltonian
\\
\begin{equation}\label{stirap-dy2}
\mathscr{H}_{eff}(z)
    = 
    \begin{pmatrix}
        0 & \kappa'_{1}(z) & 0\\
        \kappa'_{1}(z) & \Delta'(z) & \kappa'_{2}(z)\\
        0 & \kappa'_{2}(z) & 0
    \end{pmatrix}.
\end{equation}
\\
with the corresponding coefficients
\\
\begin{widetext}
\begin{subequations}
\begin{align}
    \kappa'_{1}(z) &= \kappa_{0}\,e^{-\left[az - \frac{L}{2\pi a}(a-1)\sin{\left(\frac{2\pi a}{L}z\right)} - L/2 - d\right]^{2}/s^{2}}\left[a-(a-1)\cos{\left(\frac{2\pi a}{L}z\right)}\right],
    \label{k1_new}\\
     \kappa'_{2}(z) &= \kappa_{0}\,e^{-\left[az - \frac{L}{2\pi a}(a-1)\sin{\left(\frac{2\pi a}{L}z\right)} - L/2 + d\right]^{2}/s^{2}}\left[a-(a-1)\cos{\left(\frac{2\pi a}{L}z\right)}\right],
    \label{k2_new}\\
        \Delta'(z) &= \Delta\left[a-(a-1)\cos{\left(\frac{2\pi a}{L}z\right)}\right],
    \label{k3_new}
\end{align}
\end{subequations}
\end{widetext}

in eq. \eqref{dynamics} and solve it numerically again, obtaining the results shown in Fig. \ref{new} for different values of $a$. Note, for example, that in Fig. \ref{new} d. the same full conversion process is achieved in about $8$ mm, shorter than previous realizations of a SFG cascaded protocol in bulk crystals \cite{Li,rangelov}. To do so, the new coupling coefficients must have a larger peak amplitude and narrower widths, which can be achieved by engineering the nonlinear susceptibility of the medium, but also by tuning the temperature of the crystal or  changing the pump power \cite{billat}. A recently work also seems to indicate that manipulating the transverse spatial structure of the pump field can offer a tailoring effect similar to that of engineering susceptibility directly on the crystal \cite{rafa}. 
\\

Although there seems to be no theoretical limitation on the value of the contraction parameter $a$ \cite{edson}, practical limitations will include the availability of increasingly stronger pump beams and the deleterious effects of these strong fields on the nonlinear material which can cause losses due enhanced absorption, photorefractive effect, thermal lensing, etc. On the other hand, a shorter nonlinear medium prevents large scattering effects, since 
\\
\begin{equation}
    S(L) = 1 - e^{-\alpha_s L},
\end{equation}
\\
$S(L)$ being the fraction of non-scattered photons at $L$ and $\alpha_{s}$ the scattering coefficient of the material \cite{marakis}. It also allows for more flexibility in performing a non-collinear process, since the non-collinear phase-matching angle $\theta_{nc} \propto 1/L$. 

\section{Simplification of the Coupling Coefficients}

The change in the coupling coefficients given by eqs. \eqref{k1_new}-\eqref{k3_new} can be cumbersome to achieve due to the complicated modulation given by the rescaling function. In this section we show how to choose simpler optimized coefficients and evaluate whether they can generate similar results. 

\hfill

The coupling coefficients in eqs. \eqref{k1_new} and \eqref{k2_new} represent periodic Gaussian functions modulated by an envelope given by the derivative of the rescaling function, which we can re-write as
\\
\begin{equation}\label{simple}
      g(z) = M(z)e^{-u(z)^2/s^2}.  
\end{equation}
\\
Focusing first on the Gaussian term, the peak occurs at a new coordinate $z_0$ such that $u(z_0) = 0$. Writing $z_0 = Lx_0/2\pi a$ and comparing $u(z)$ and eq. \eqref{simple} with eqs. \eqref{k1_new} and \eqref{k2_new} leads to the transcendental equation 
\\
\begin{equation}
    ax_0 - (a-1)\sin(x_0) = \pi a \pm\frac{\pi a}{5},
\end{equation}
\\
which we can solve for each value of $a$ to determine $z_0$. The $\pm \pi a/5$ factor is used whether $\mp d$ appears in the original coefficient. Expanding $u(z)$ around $z_0$ and keeping up to the second term then yields 
\\
\begin{equation}\label{gaus}
    g(z) \propto e^{-(z-z_0)^2/s_{new}^2},
\end{equation}
\\
where $s_{new} = s/|u'(z_0)|$, $u'(z_0)$ being the derivative of $u(z)$ with respect to $z$ at $z=z_0$.

\hfill

If the envelope $M(z)$ oscillates slowly compared to the Gaussian drop around $z_0$, we can use the least squares fit to substitute this function by an optimized constant $M_c$ such that  \cite{blais}
\\
\begin{equation}
  \frac{dE}{dM_c} = 0,
\end{equation}
\\
where $E$ is the error function
\\
\begin{equation}
    E = \int [\kappa'_{1,2}(z) - M_c\,g(z)]^2 \, dz,
\end{equation}
\\
and which, to our case, has the analytic solution
\\
\begin{equation}\label{Mc}
    M_c = a - (a-1)\cos\left(\frac{2\pi a z_0}{L}\right)e^{-\pi^2a^2s^2_{new}/2L^2}.
\end{equation}
\\
Combining eqs. \eqref{gaus} and \eqref{Mc} therefore suggests to use the approximate coefficients
\\
\begin{subequations}
\begin{align}
    \kappa''_{1}(z) &= (\kappa_{0}M_c)\,e^{-(z - z_{01})^2/s^2_{new}},
    \label{k1_simp}\\
     \kappa''_{2}(z) &= (\kappa_{0}M_c)\,e^{-(z - z_{02})^2/s^2_{new}},
    \label{k2_simp}\\
        \Delta'(z) &= \Delta M_c.
    \label{k3_simp}
\end{align}
\end{subequations}

\hfill

\begin{figure}[h!]
    \centering
    \includegraphics[width=0.80\linewidth]{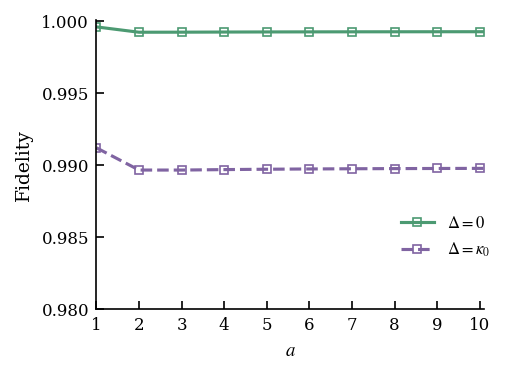}
    \caption{Fidelity under simplified conditions. Using the simplified expressions for the coupling coefficients and detuning presented in eqs. \eqref{k1_simp}-\eqref{k3_simp}, we calculate the Fidelity of the protocol in the phase-matched and detuned situations. The results indicate that the protocol remains robust under these circumstances.}
    \label{simp}
\end{figure}

Using the definition of fidelity as 
\\
\begin{equation}
    F = |\braket{3|\psi_f}|^{2}
\end{equation}
\\
we again solve eq. \eqref{dynamics} numerically and evaluate the quality of the new coefficients by calculating the fidelity for different values of the contraction parameter $a$, as shown in Fig. \ref{simp}, where the green curve represents a phase-matched situation and the purple curve shows the behavior with mismatch. In both cases, we still attained a high fidelity of $F > 0.985$. 

\section{Application in Single Photon Detection}

In order to illustrate the concrete usefulness of the method, we consider the practical situation of detecting single photons in the mid-infrared (MIR) region of 3-5 $\mu$m. This is an important bandwidth, especially for spectroscopy \cite{gray} and long distance quantum key distribution in free space \cite{ding}, but challenging for single photon detection because it requires semiconductors with narrow bandgaps and due to the presence of considerable thermal noise \cite{Lau,russo}. An interesting possibility becomes up-converting these photons into a bandwidth where they can be more easily measured, for example, within the $400-800$ nm range, where silicon-based single-photon avalanche detectors (SPADs) can reach efficiencies greater than $0.9$. 

\hfill

The protocol discussed in the previous sections could, in principle, be applied to create a compact frequency conversion-based single photon detector that supplies the needs of the 3-5 $\mu$m region. To do so, we set the pump field in 1.55 $\mu$m, a wavelength where short linewidth, low noise DFB lasers are easily available with typical power outputs in the order of $10^{-1}$ W, amplified up to the order of $10^{1}$ W by commercially available Erbium-doped fiber amplifiers (EDFAs). Even larger amplification factors can be found by setting the whole system inside a resonating cavity. The nonlinear medium chosen is a $14$ mm MgO-doped Lithium Niobate (LN) crystal with the required double-Gaussian polling pattern, which can be custom-made by several companies that offer aperiodically poled crystals for quantum technology applications. MgO-LiNbO$_3$ is a particularly good choice due to the low linear and two-photon absorption around 1.55 $\mu$m and negligible photorefractive effects \cite{hcp}.

\hfill

As an example, we take an incoming photon with a wavelength of 4 $\mu$m (right in the middle of the bandwidth) and let it interact with the strong pump field inside the crystal. The cascaded process will then convert 4 $\mu$m $\rightarrow$ 0.65 $\mu $m, without significantly populating the intermediate frequency at 1.12 $\mu$m. For the original STIRAP process, using a 70 mm crystal, we estimate from eq. \eqref{ad-con} a minimum coupling coefficient of about $0.15$ cm$^{-1}$, which would correspond to a pump power of about $1$ W for a loosely focused beam with waist of 100 $\mu$m. The rescaled process on the 14 mm crystal would then involve a $a=5$ parameter, meaning that the peak coupling coefficient would be, according to eq. \eqref{Mc}, approximately $\kappa_0 = 1.27 $cm$^{-1}$, corresponding to an increased pump power of $\sim$ 50 W, which is still below the damage threshold of the material if it has at least $5\%$ of MgO doping. 

\hfill

\begin{figure}[h!]
    \centering
    \includegraphics[width=0.80\linewidth]{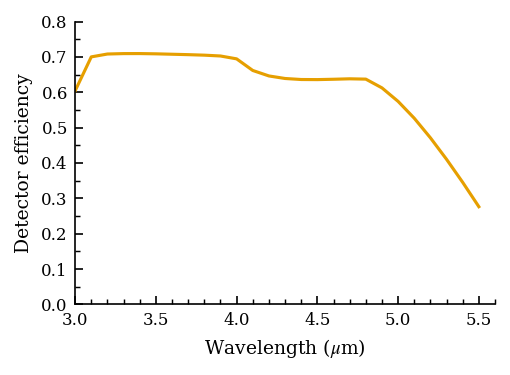}
    \caption{Detector efficiency on the $3-5\mu$m bandwidth. Using a good silicon-based SPAD with detection efficiency of about 0.9, the cascaded frequency conversion approach seems to lead to theoretical efficiencies of up to 0.75.}
    \label{last}
\end{figure}

In Fig. \ref{last} we plot the efficiency of this proposed detector, considering the transmission window of Lithium Niobate in the 3-5 $\mu$m region and a SPAD detection efficiency for the up-converted photon of about 0.9. This result seems to indicate a reasonably high fidelity for our theoretical device  when compared to other non-cryogenic detection devices like those shown in \cite{russo}.

\section{Discussion and Conclusion} 

In this work we showed how the coupled equations for the cascaded conversion process can be described analogously to the dynamics of the STIRAP protocol. This method is robust in promoting complete frequency conversion between photons from a seed field $\omega_s$ to an up-converted field $\omega_{u}$, using a strong pump field $\omega_p$ without significantly populating the intermediary field $\omega_i$. The trade-off to this protocol is that, since it is adiabatic, it requires longer nonlinear materials which can cause optical losses and create difficulties for the integration of the converter in small photonic devices. This situation can be improved by proposing a shortcut to adiabaticity to the STIRAP, using the time rescaling method. By properly shaping the coupling coefficients, we showed that full conversion can still be achieved at a fraction of the original medium length. 

It is worth noticing that, as shown in \cite{lukas}, the TR method is such that the system follows the same path in the Hilbert space as the original protocol, instead of simply sharing the same initial and final states. Therefore, it inherits from the original protocol not only full conversion, but also its properties such as robustness against errors in the coupling coefficients, increase of detuning, etc. This robustness in turn make the protocol less sensitive to parameters such as crystal temperature or sharp bandwidths.

In section \ref{stirap}, the crystal length was set as $L = 60$ mm as a reasonable value to maintain the system within the adiabatic regime, following eq. \eqref{ad-con}. This choice is not in any way a hard limit on the size of current frequency converters. In practice, there is a plethora of possible sizes for conversion devices, depending on their underlying physical phenomena, from several centimeters associated to bulk crystals and waveguides \cite{Barros,pan} down to the sub-millimeter scale for microresonators \cite{Cao}. The possible benefits and challenges of applying our proposal to each particular configuration must be evaluated in a case-by-case basis.

\hfill

Since shaping the coefficients with the modulation functions required by the TR method can be difficult, we evaluated in section 4 how the conversion process would work if these coefficients were approximated by optimized Gaussian functions. The result of our simulations shows high fidelity in frequency conversion, both in the phase-matched and phase-mismatched scenarios. This last result helps to simplify future experimental implementations of the protocol and is applicable not only to frequency conversion but also to several systems to which the TR technique considered, therefore presenting a useful novelty for the method itself. 

Finally, in section 5 we proposed a possible application of the TR cascaded frequency conversion in the designing of single photon detectors in the MIR region. Using commercially available components, our theoretical results indicate good efficiency within the scope of non-cryogenic detectors. This proposal does not intend to be a hard blueprint on the construction of single photon detectors, but an illustration of the feasibility of the protocol and an inspiration for future inquiries and research on the method. Our results also offer possible insight on the use of optical systems as experimental benchmarks to test the strengths and limitations of the time-rescaling shortcut technique itself. The investigation of the use of the protocol introduced here to promote cascaded conversion of strong classical fields in order to test the boundaries of a STIRAP-like dynamics, will be the subject of future inquiries. 

\section*{Funding}
This work was supported by FAPESP (São Paulo Research Foundation) under Grant No. 2025/00093-6.

\section*{Disclosures}
The author declares no conflicts of interest.

\section*{Data availability}
Data underlying the results presented in this paper are openly available on Figshare at \url{https://doi.org/10.6084/m9.figshare.34059534}.

\bibliography{apssamp}

\end{document}